\documentclass[12pt]{iopart}

\usepackage{cite}
\usepackage{algorithmic}
\usepackage{graphicx}
\usepackage{textcomp}
\usepackage{xcolor}

\usepackage{tikz}
\usepackage{pgfplots}
\usetikzlibrary{arrows,shapes,circuits.logic.US,shapes.gates.logic.IEC,calc,decorations.markings,patterns,positioning,fit,arrows.meta}

\usepackage{subcaption}
\usepackage{threeparttable}
\usepackage{multirow,multicol}

\usepackage{url}

\newcommand{\blank}{\vspace{0cm}}

\begin{document}

\title[Journal of Neural Engineering]{E-Sort: Empowering End-to-end Neural Network for Multi-channel Spike Sorting with Transfer Learning and Fast Post-processing}

\author{Yuntao Han, Shiwei Wang}

\address{Institute for Integrated Micro and Nano Systems, School of Engineering, University of Edinburgh, Edinburgh, UK}
\ead{\{Yuntao.Han, shiwei.wang\}@ed.ac.uk}
\vspace{10pt}
\begin{indented}
\item[]December 2024
\end{indented}

\begin{abstract}
Decoding extracellular recordings is a crucial task in electrophysiology and brain-computer interfaces. Spike sorting, which distinguishes spikes and their putative neurons from extracellular recordings, becomes computationally demanding with the increasing number of channels in modern neural probes. To address the intensive workload and complex neuron interactions, we propose E-Sort, an end-to-end neural network-based spike sorter with transfer learning and parallelizable post-processing. Our framework reduces the required number of annotated spikes for training by 44\% compared to training from scratch, achieving up to 25.68\% higher accuracy. Additionally, our novel post-processing algorithm is compatible with deep learning frameworks, making E-Sort significantly faster than state-of-the-art spike sorters. On synthesized Neuropixels recordings, E-Sort achieves comparable accuracy with Kilosort4 while sorting 50 seconds of data in only 1.32 seconds. Our method demonstrates robustness across various probe geometries, noise levels, and drift conditions, offering a substantial improvement in both accuracy and runtime efficiency compared to existing spike sorters.
\end{abstract}

%
\vspace{2pc}
\submitto{\JNE}
%
%
%

\section{Introduction}

Decoding extracellular recordings is an essential task for both electrophysiological studies and brain-computer interface (BCI) applications~\cite{zhang2023spike}. Spike sorting is the first step in distinguishing spikes and their putative neurons from these recordings~\cite{ss}. Conventional spike sorters comprise several steps, i.e., spike detection, feature extraction, and clustering.

Recently, the advancement of neural probes enables the recording of thousands of tightly-placed electrodes. For example, the Neuropixels (NP) series (1.0~\cite{np1}, 2.0~\cite{np2}, Ultra~\cite{npultra}) are capable of recording $384$ channels in parallel with electrode pitches of $20$, $15$, $6$ um, respectively. The large number of channels causes an increase in computational workload and processing time, while the shrinkage of pitches causes higher noise levels~\cite{npultra} and more drifting problems~\cite{survey_hd}, compromising the accuracy of the spike sorters.

Modern spike sorters~\cite{ks4}\cite{dartsort} utilized deconvolutions and drift registrations iteratively in time chunks to counteract the accuracy degradation. However, these techniques required more computations and struggled in noisy environments and complex neuron distributions.

With the neural networks (NNs) demonstrating promising performance in various fields, several studies on NN-assisted spike sorters have been conducted to replace some steps in conventional spike sorting pipelines~\cite{yass}\cite{lts}. For example, YASS~\cite{yass} used two NNs for spike detection and waveform denoising, respectively, each of which was constructed with convolutions on the temporal and spatial dimensions. However, because only specific steps in the spike sorting pipelines were substituted with NNs, other non-NN processing algorithms were still present, which may require manual tuning when adapting to different recordings. There were also NN-only designs utilizing several NNs~\cite{multinn} or branches~\cite{elvisort} for specific stages in spike sorting. Nevertheless, these designs were difficult to optimize for computational efficiency, since several different algorithms were involved which are hard to ameliorate collectively.

By contrast, the solutions based on end-to-end NNs, which bind the whole spike sorting pipeline into a single NN, are promising because of their simplicity and integrity. DualSort~\cite{dualsort} demonstrated an end-to-end NN for sorting a single-channel recording by incorporating data augmentation based on temporal shifting and population-based post-processing. However, there were several concerns about adapting this design to the recordings from high-channel-count neural probes.

\textbf{Manual annotations of neurons:} Because the number of output neurons needs to be determined case by case, each model should be trained separately for each recording. Since the NN training requires several annotated spikes for each neuron, the total number of annotated spikes increases with the number of neurons. Moreover, the number of annotated spikes for each neuron should also be increased to achieve accurate sorting because of the additional spatial pattern compared to single-channel recordings. While few-shot training with adversarial representation learning has been used in a previous study~\cite{lts}, it was limited to single-channel recordings and only performed feature extraction on the neural network with only tens of spikes utilized during training.

\textbf{Non-parallelizable post-processing:} Though the NN accelerations are well supported by the hardware for parallel computation (e.g., GPUs), the post-processing in DualSort is difficult in parallelization because of the data dependency in the temporal dimension, i.e., the output of a specific timestep is dependent on its former timestep. Therefore, despite the NN inference could be computed in parallel, the post-processing has to be performed one step at a time, leading to excessive execution time.
 
To tackle the aforementioned challenges, we propose E-Sort\footnote{Our code is available at https://github.com/LucasHanCEF/e-sort .}, a novel spike sorter utilizing an end-to-end NN trained with transfer learning and a parallelizable post-processing scheme. Our contributions are as follows:

\begin{itemize}
    \item We propose a framework of transfer learning for end-to-end spike sorting. We construct a neural network with temporal and spatial filters, similar to~\cite{yass}. We pre-train these filters on large recordings and fine-tune these filters, along with training the classifier from scratch with the starting few spikes from the test recording.
    \item We design a post-processing composed of a triangle filter followed by peak detection and thresholding for rejecting redundant spikes and suppressing false positives (noises). This post-processing is GPU-friendly and compatible with NN frameworks.
    \item We evaluate E-Sort on synthesized recordings. Compared with the state-of-art spike sorter for Neuropixels, Kilosort4~\cite{ks4}, our proposed strategy for transfer learning achieves comparable accuracy with $2.25\times$ reduction on the duration of the training set compared with training from scratch. Moreover, our design significantly reduces execution time compared with all existing spike sorters while achieving an accuracy comparable with Kilosort4, a mainstream spike sorting algorithm commonly used in the neuroscience community.
\end{itemize}

\begin{figure}[t]
\captionsetup[subfigure]{labelformat=parens, labelsep=period, justification=raggedright, singlelinecheck=false}

\begin{subfigure}[b]{\linewidth}
  \centering
  \caption{}
  \vspace{-10pt}
  \includegraphics[width=.95\linewidth]{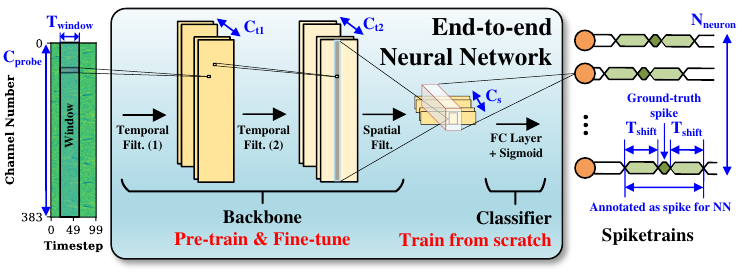}
  \label{fig:nn}
\end{subfigure}

\begin{subfigure}[b]{\linewidth}
  \centering
  \caption{}
  \vspace{-10pt}
  \includegraphics[width=.9\linewidth]{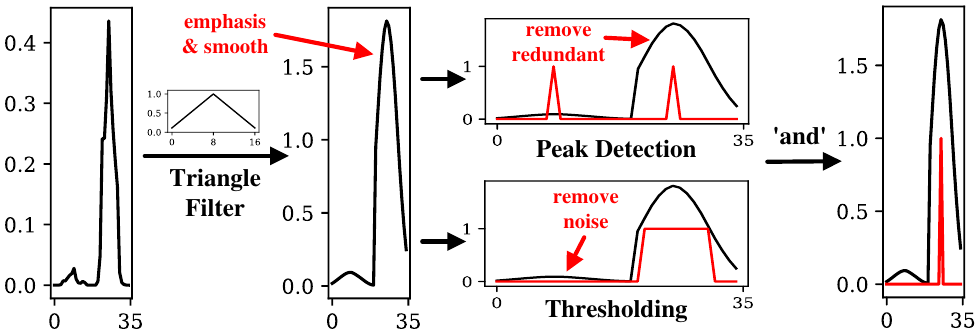}
  \label{fig:post-processing}
\end{subfigure}
\caption{The proposed E-Sort spike sorting framework. (a) End-to-end NN pre-training \& finetuning scheme. (b) Parallelizable post-processing for removing noises and redundant spikes.}
\label{fig:fafesort}
\blank
\end{figure}

\section{Dataset}

In this letter, we use the synthesized recordings for training and examining the proposed framework. The synthesized recordings are generated by the MEARec~\cite{mearec} package, which includes $13$ layer $5$ juvenile rat somatosensory cortex neuron models from the neocortical microcircuit collaboration portal~\cite{portal}. This package allows the customization of recordings with different characteristics, e.g., probe geometry, noise level and number of included neurons. The synthesized recordings come with the ground truth spiketrains, facilitating the supervised training and evaluations of the proposed framework.

\section{Methodology}

\subsection{Dataset Preparation}

To utilize both the spatial and temporal information involved in the recording, we define the windows involving all probe channels $C_{\emph{probe}}$ within a time interval $T_{\emph{window}}$ as the input for the NN. However, the spikes in the extracellular recordings are not present for the most time in the recording and the firing rates of different neurons vary significantly, e.g., excitatory cells fire several times faster than inhibitory cells. Using all windows to build the dataset would result in imbalances in both spike/non-spike ratio and spikes from different neurons, which will diminish the performance of the NN in identifying the sparse spikes from the recording and distinguishing the putative neurons for each spike, respectively.

To increase the numbers of the spike-present windows and balance the number of windows with spikes from different putative neurons in the dataset, we establish the dataset in three steps as follows: firstly, all windows with at least one spike in the centre are selected; secondly, the selected windows are augmented allowing a temporal shift $T_{\emph{shift}}$ with paddings from the recording, and therefore each spike is picked up by $2T_{\emph{shift}}+1$ windows, increasing the number of spike-present samples in the dataset; finally, an equal number of non-spike windows is randomly selected from the recording, ensuring that the total number of non-spike windows matches the number of spike-present windows, thus balancing the number of spike/non-spike samples in the dataset.

\subsection{Transfer Learning}

Transfer learning refers to the learning technique for pre-training a neural network on a large dataset and fine-tuning for a new small dataset. In this paper, we utilize this technology to decrease the number of annotated spikes required for acquiring a high-accuracy NN for a specific recording.

The NN architecture comprises of the backbone and the classifier for feature extraction and classification respectively, as shown in Fig.~\ref{fig:nn}. The backbone is constructed with two temporal filters with channel count $C_{t1}$ and $C_{t2}$ and one spatial filter with channel count $C_{s}$, similar to what is used in YASS~\cite{yass}. These filters extract the latent features from the temporal and spatial dimensions with linear mappings on the corresponding dimensions, each of which is followed by normalization and ReLU activation for regulation and introducing nonlinearity. We pre-train the backbone on long-duration and neuron-rich recordings to find general features across different classes of spikes. The classifier featuring a fully-connected layer and a softmax layer, is trained from scratch for each recording because the number of output channels is dependent on the number of neurons in the recording.

\subsection{Post-processing}

Because of the temporal shifting applied for augmenting the dataset, each spike should be detected repeatedly by the NN. To remove redundant spikes and to avoid false positive detection (e.g., noises mis-detected as spikes), we propose a post-processing framework, which is parallelizable and therefore compatible with modern deep learning framework, as shown in Fig.~\ref{fig:post-processing}.

Considering that the noise is unlikely to be detected continuously in the temporal dimension, and each spike should be detected continuously, we apply a triangle filter on the temporal dimension to emphasise and smooth the output from the NN. The emphasized spiketrain is subsequently processed by two algorithms. The first one is the peak detection for removing the redundant spikes, which finds local maxima by comparing each data point with two temporally adjacent points. The second one is the thresholding process for removing the noises, which filters out the data points that exceed a certain threshold. The final detected spikes are calculated as the data points satisfying both conditions.

\section{Experiments}
\label{sec:experiments}

\subsection{Experimental Setup}

The synthesized recordings are generated using MEARec with a fixed duration of 100 seconds. The training set is sampled from the first 50 seconds, while the last 50 seconds are used for testing. Different random seeds are applied to all processes to generate separate recordings for pre-training and fine-tuning. The neuron densities are set to 1.5e4 and 1e4 per mm$^3$ in pre-training and fine-tuning recordings, respectively, and $T_{\emph{shift}}$ is set to $5$. The proposed E-Sort is implemented using the PyTorch framework. All training is performed in 50 epochs under the Adam optimizers with a learning rate of 5e-3 for rapid convergence. The experiment platform is equipped with 32 CPU cores from the Intel Xeon Platinum 8468 CPU with 128GB physical memory and an Nvidia H100 GPU.

\subsection{Transfer Learning Results}

\begin{figure}
\centering

\begin{tikzpicture}
\begin{axis}[
    width=0.9\linewidth,
    height=.5\linewidth,
    xlabel={\#Spikes per neuron},
    ylabel={Accuracy ($\%$)},
    ylabel near ticks,
    xmin=3, xmax=37,
    xtick={5,10,15,20,25,30,35},
    x tick label style={rotate=60, anchor=east},
    grid=major,
    legend entries={Scratch,Pretrained},
    legend pos=south east,
    legend cell align={left}
]


\addplot[gray,mark=*,mark size=1pt,line width=1pt] coordinates {
(3, 48.83)
(4, 48.64)
(5, 64.93)
(6, 68.35)
(7, 70.17)
(8, 73.07)
(9, 78.14)
(10, 79.16)
(11, 78.68)
(12, 81.62)
(13, 83.11)
(14, 84.88)
(15, 85.48)
(16, 82.98)
(17, 84.09)
(18, 87.21)
(19, 85.62)
(20, 85.17)
(21, 88.04)
(22, 87.61)
(23, 88.27)
(24, 89.00)
(25, 89.18)
(26, 87.45)
(27, 88.84)
(28, 87.41)
(29, 88.95)
(30, 90.76)
(31, 91.03)
(32, 90.45)
(33, 90.76)
(34, 89.92)
(35, 90.46)
(36, 91.42)
(37, 90.19)
}; \label{line:scratch}

\addplot[black,mark=*,mark size=1pt,line width=1pt] coordinates {
(3, 63.56)
(4, 74.32)
(5, 78.73)
(6, 81.86)
(7, 84.06)
(8, 85.87)
(9, 86.95)
(10, 87.99)
(11, 88.34)
(12, 89.37)
(13, 90.15)
(14, 90.69)
(15, 90.98)
(16, 91.32)
(17, 91.75)
(18, 91.89)
(19, 92.16)
(20, 92.22)
(21, 92.45)
(22, 92.70)
(23, 92.69)
(24, 93.17)
(25, 93.24)
(26, 93.20)
(27, 93.31)
(28, 92.53)
(29, 93.64)
(30, 93.73)
(31, 93.56)
(32, 93.78)
(33, 93.79)
(34, 93.76)
(35, 93.64)
(36, 94.11)
(37, 94.05)
}; \label{line:finetune}

\draw[dashed,ultra thick,blue] (axis cs:3, 91.58) -- (axis cs:37, 91.58);
\node at (axis cs:3, 90.5) [anchor=south west] {\color{blue}\textbf{91.58\%@KS4}};
\draw[ultra thick,blue] (axis cs:36, 74) -- (axis cs:36, 91.42);
\draw[ultra thick,blue] (axis cs:16, 74) -- (axis cs:16, 91.32);
\draw[-{Latex}, ultra thick,blue] (axis cs:36, 81) -- (axis cs:16, 81);
\node at (axis cs:26, 81) [anchor=north] {\color{blue}$\mathbf{2.25\times}$ \textbf{reduction}};

\draw[dashed,ultra thick,blue] (axis cs:4, 48.64) -- (axis cs:4, 74.32);
\draw[ultra thick,blue] (axis cs:4, 48.64) -- (axis cs:12, 48.64);
\draw[ultra thick,blue] (axis cs:4, 74.32) -- (axis cs:12, 74.32);
\draw[-{Latex}, ultra thick,blue] (axis cs:10, 48.64) -- (axis cs:10, 74.32);
\node at (axis cs:10, 61.48) [anchor=west] {\color{blue}\shortstack{$\mathbf{+25.68\%}$\\\textbf{\ increment}}};

\end{axis}
\end{tikzpicture}
\caption{
Achieved accuracy versus number of spikes per neuron for finetuning pretrained models and training models from scratch.
}
\label{fig:few_shot_comprison}
\blank
\end{figure}
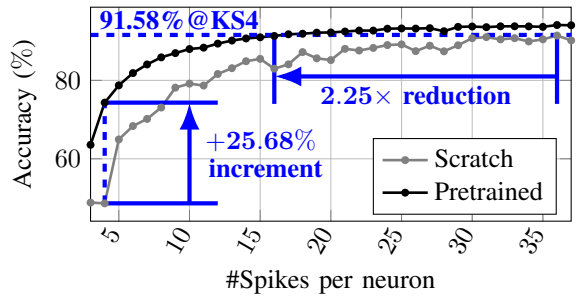

To evaluate the performance of the pre-trained model on the synthesized recordings, the pre-trained model is acquired by training with all spikes from the first 50 seconds of the recording for pre-training. We then assess the accuracy achieved with various number of spikes per neuron for fine-tuning, which is denoted as $N_{\emph{ft}}$, as shown in Fig.~\ref{fig:few_shot_comprison}. The accuracy is defined as:

\begin{equation}
    \emph{accuracy} = \frac{\#TP}{\#TP+\#FP+\#FN}
\end{equation}
where $\#TP$, $\#FP$, and $\#FN$ are the number of true positive (spikes sorted correctly), false positive (noises mis-sorted as spikes), and false negative (spikes mis-sorted as noises) results.

Compared to the model trained from scratch, fine-tuning from the pre-trained model achieves a higher accuracy of up to $25.68\%$ improvement. This difference decreases when more spikes are used for fine-tuning/training because the model trained from scratch progresses towards learning features from the increased samples, which diminishes the significance of the pre-learned features from pre-training. To achieve a comparable accuracy (less than $0.5\%$ difference) with Kilosort4 (KS4), the pre-trained model requires $16$ spikes per neuron for fine-tuning, while the model trained from scratch requires $36$ spikes per neuron, indicating a $2.25\times$ reduction in the training set.

\subsection{Generality of pre-trained Model}

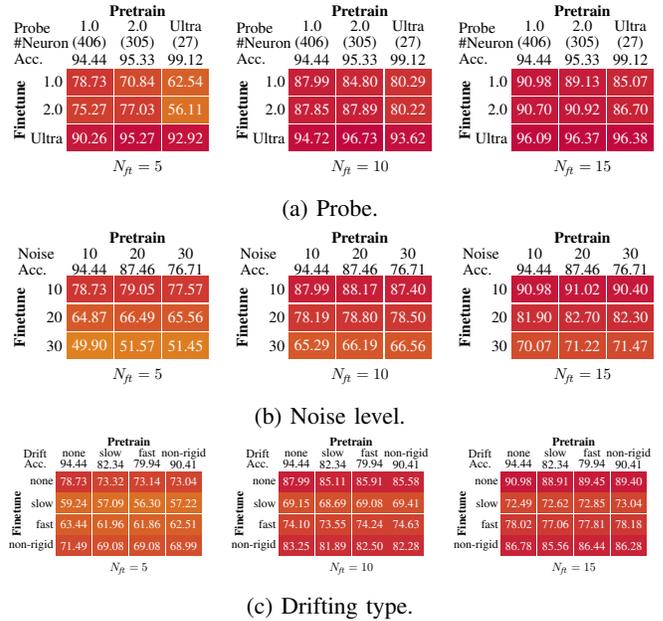
\begin{figure}[t]
\centering
\begin{subfigure}[b]{\linewidth}
\centering

\resizebox{.32\linewidth}{!}{
\begin{tikzpicture}[scale=0.6]

    \node[minimum size=6mm] at (3.6, 2.1) {\textbf{Pretrain}};
    \node[minimum size=6mm, rotate=90, anchor=south] at (-1, -2.5) {\textbf{Finetune}};

    \node[minimum size=6mm] at (-.5, .6) {\shortstack[l]{Probe\\\#Neuron\\Acc.}};

    \node[minimum size=6mm] at (2.05, .6) {\shortstack{1.0\\(406)\\94.44}};
    \node[minimum size=6mm] at (4.1, .6) {\shortstack{2.0\\(305)\\95.33}};
    \node[minimum size=6mm] at (6.15, .6) {\shortstack{Ultra\\(27)\\99.12}};

  \foreach \y [count=\n] in {
      {78.73,70.84,62.54},
      {75.27,77.03,56.11},
      {90.26,95.27,92.92},
    } {
      \foreach \x [count=\m] in \y {
        \node[fill=purple!\x!yellow, minimum size=6mm,text width=9mm, text=white, align=center] at (2.05*\m,-1.05*\n) {\x};
      }
    }

  \foreach \a [count=\i] in {1.0,2.0,Ultra} {
    \node[minimum size=6mm, anchor=east] at (1,-1.05*\i) {\a};
  }

  \node[minimum size=6mm] at (3.6, -4.3) {$N_{\emph ft}=5$};
  
\end{tikzpicture}
}
\resizebox{.32\linewidth}{!}{
\begin{tikzpicture}[scale=0.6]

    \node[minimum size=6mm] at (3.6, 2.1) {\textbf{Pretrain}};
    \node[minimum size=6mm, rotate=90, anchor=south] at (-1, -2.5) {\textbf{Finetune}};

    \node[minimum size=6mm] at (-.5, .6) {\shortstack[l]{Probe\\\#Neuron\\Acc.}};

    \node[minimum size=6mm] at (2.05,.6) {\shortstack{1.0\\(406)\\94.44}};
    \node[minimum size=6mm] at (4.1, .6) {\shortstack{2.0\\(305)\\95.33}};
    \node[minimum size=6mm] at (6.15,.6) {\shortstack{Ultra\\(27)\\99.12}};

  \foreach \y [count=\n] in {
      {87.99,84.80,80.29},
      {87.85,87.89,80.22},
      {94.72,96.73,93.62},
    } {
      \foreach \x [count=\m] in \y {
        \node[fill=purple!\x!yellow, minimum size=6mm,text width=9mm, text=white, align=center] at (2.05*\m,-1.05*\n) {\x};
      }
    }

  \foreach \a [count=\i] in {1.0,2.0,Ultra} {
    \node[minimum size=6mm, anchor=east] at (1,-1.05*\i) {\a};
  }

  \node[minimum size=6mm] at (3.6, -4.3) {$N_{\emph ft}=10$};
  
\end{tikzpicture}
}
\resizebox{.32\linewidth}{!}{
\begin{tikzpicture}[scale=0.6]

    \node[minimum size=6mm] at (3.6, 2.1) {\textbf{Pretrain}};
    \node[minimum size=6mm, rotate=90, anchor=south] at (-1, -2.5) {\textbf{Finetune}};

    \node[minimum size=6mm] at (-.5, .6) {\shortstack[l]{Probe\\\#Neuron\\Acc.}};

    \node[minimum size=6mm] at (2.05,.6) {\shortstack{1.0\\(406)\\94.44}};
    \node[minimum size=6mm] at (4.1, .6) {\shortstack{2.0\\(305)\\95.33}};
    \node[minimum size=6mm] at (6.15,.6) {\shortstack{Ultra\\(27)\\99.12}};

  \foreach \y [count=\n] in {
      {90.98,89.13,85.07},
      {90.70,90.92,86.70},
      {96.09,96.37,96.38},
    } {
      \foreach \x [count=\m] in \y {
        \node[fill=purple!\x!yellow, minimum size=6mm,text width=9mm, text=white, align=center] at (2.05*\m,-1.05*\n) {\x};
      }
    }

  \foreach \a [count=\i] in {1.0,2.0,Ultra} {
    \node[minimum size=6mm, anchor=east] at (1,-1.05*\i) {\a};
  }

  \node[minimum size=6mm] at (3.6, -4.3) {$N_{\emph ft}=15$};
  
\end{tikzpicture}
}
\caption{Probe.}
\label{fig:probe}
\end{subfigure}
\begin{subfigure}[b]{\linewidth}
\centering

\resizebox{.32\linewidth}{!}{
\begin{tikzpicture}[scale=0.6]

    \node[minimum size=6mm] at (3.6, 1.3) {\textbf{Pretrain}};
    \node[minimum size=6mm, rotate=90, anchor=south] at (-.5, -2.4) {\textbf{Finetune}};

    \node[minimum size=6mm] at (-.5, .2) {\shortstack[l]{Probe\\Acc.}};

    \node[minimum size=6mm] at (2.05,.2) {\shortstack{10\\94.44}};
    \node[minimum size=6mm] at (4.1, .2) {\shortstack{20\\87.46}};
    \node[minimum size=6mm] at (6.05,.2) {\shortstack{30\\76.71}};

  \foreach \y [count=\n] in {
      {78.73,79.05,77.57},
      {64.87,66.49,65.56},
      {49.90,51.57,51.45},
    } {
      \foreach \x [count=\m] in \y {
        \node[fill=purple!\x!yellow, minimum size=6mm,text width=9mm, text=white, align=center] at (2.05*\m,-1.05*\n) {\x};
      }
    }

  \foreach \a [count=\i] in {10,20,30} {
    \node[minimum size=6mm, anchor=east] at (1,-1.05*\i) {\a};
  }

  \node[minimum size=6mm] at (3.6, -4.3) {$N_{\emph ft}=5$};
  
\end{tikzpicture}
}
\resizebox{.32\linewidth}{!}{
\begin{tikzpicture}[scale=0.6]

    \node[minimum size=6mm] at (3.6, 1.3) {\textbf{Pretrain}};
    \node[minimum size=6mm, rotate=90, anchor=south] at (-.5, -2.4) {\textbf{Finetune}};

    \node[minimum size=6mm] at (-.5, .2) {\shortstack[l]{Probe\\Acc.}};

    \node[minimum size=6mm] at (2.05,.2) {\shortstack{10\\94.44}};
    \node[minimum size=6mm] at (4.1, .2) {\shortstack{20\\87.46}};
    \node[minimum size=6mm] at (6.15,.2) {\shortstack{30\\76.71}};

  \foreach \y [count=\n] in {
      {87.99,88.17,87.40},
      {78.19,78.80,78.50},
      {65.29,66.19,66.56},
    } {
      \foreach \x [count=\m] in \y {
        \node[fill=purple!\x!yellow, minimum size=6mm,text width=9mm, text=white, align=center] at (2.05*\m,-1.05*\n) {\x};
      }
    }

  \foreach \a [count=\i] in {10,20,30} {
    \node[minimum size=6mm, anchor=east] at (1,-1.05*\i) {\a};
  }

  \node[minimum size=6mm] at (3.6, -4.3) {$N_{\emph ft}=10$};
  
\end{tikzpicture}
}
\resizebox{.32\linewidth}{!}{
\begin{tikzpicture}[scale=0.6]

    \node[minimum size=6mm] at (3.6, 1.3) {\textbf{Pretrain}};
    \node[minimum size=6mm, rotate=90, anchor=south] at (-.5, -2.4) {\textbf{Finetune}};

    \node[minimum size=6mm] at (-.5, .2) {\shortstack[l]{Probe\\Acc.}};

    \node[minimum size=6mm] at (2.05,.2) {\shortstack{10\\94.44}};
    \node[minimum size=6mm] at (4.1, .2) {\shortstack{20\\87.46}};
    \node[minimum size=6mm] at (6.15,.2) {\shortstack{30\\76.71}};

  \foreach \y [count=\n] in {
      {90.98,91.02,90.40},
      {81.90,82.70,82.30},
      {70.07,71.22,71.47},
    } {
      \foreach \x [count=\m] in \y {
        \node[fill=purple!\x!yellow, minimum size=6mm,text width=9mm, text=white, align=center] at (2.05*\m,-1.05*\n) {\x};
      }
    }

  \foreach \a [count=\i] in {10,20,30} {
    \node[minimum size=6mm, anchor=east] at (1,-1.05*\i) {\a};
  }

  \node[minimum size=6mm] at (3.6, -4.3) {$N_{\emph ft}=15$};
  
\end{tikzpicture}
}
\caption{Noise level.}
\label{fig:noise_level}
\end{subfigure}
\begin{subfigure}[b]{\linewidth}
\centering

\resizebox{.32\linewidth}{!}{
\begin{tikzpicture}[scale=0.6]

    \node[minimum size=6mm] at (4.5, 1.1) {\textbf{Pretrain}};
    \node[minimum size=6mm, rotate=90, anchor=south] at (-.5, -2.2) {\textbf{Finetune}};

    \node[minimum size=6mm, anchor=south] at (-.1, -.7) {Acc.};
    \node[minimum size=6mm, anchor=south] at (-.1, -.17) {Drift};

    \node[minimum size=6mm, anchor=south] at (2.05, -.6) {\shortstack{none\\94.44}};
    \node[minimum size=6mm, anchor=south] at (4.1, -.6) {\shortstack{slow\\82.34}};
    \node[minimum size=6mm, anchor=south] at (6.15, -.6) {\shortstack{fast\\79.94}};
    \node[minimum size=6mm, anchor=south] at (8.2, -.7) {90.41};
    \node[minimum size=6mm, anchor=south] at (8.2, -.11) {non-rigid};

  \foreach \y [count=\n] in {
      {78.73,73.32,73.14,73.04},
      {59.24,57.09,56.30,57.22},
      {63.44,61.96,61.86,62.51},
      {71.49,69.08,69.08,68.99},
    } {
      \foreach \x [count=\m] in \y {
        \node[fill=purple!\x!yellow, minimum size=6mm,text width=9mm, text=white, align=center] at (2.05*\m,-1.05*\n) {\x};
      }
    }

  \foreach \a [count=\i] in {none,slow,fast,non-rigid} {
    \node[minimum size=6mm, anchor=east] at (1,-1.05*\i) {\a};
  }

  \node[minimum size=6mm] at (4.5, -5.35) {$N_{\emph ft}=5$};
  
\end{tikzpicture}
}
\resizebox{.32\linewidth}{!}{
\begin{tikzpicture}[scale=0.6]

    \node[minimum size=6mm] at (4.5, 1.1) {\textbf{Pretrain}};
    \node[minimum size=6mm, rotate=90, anchor=south] at (-.5, -2.2) {\textbf{Finetune}};

    \node[minimum size=6mm, anchor=south] at (-.1, -.7) {Acc.};
    \node[minimum size=6mm, anchor=south] at (-.1, -.17) {Drift};

    \node[minimum size=6mm, anchor=south] at (2.05, -.6) {\shortstack{none\\94.44}};
    \node[minimum size=6mm, anchor=south] at (4.1, -.6) {\shortstack{slow\\82.34}};
    \node[minimum size=6mm, anchor=south] at (6.15, -.6) {\shortstack{fast\\79.94}};
    \node[minimum size=6mm, anchor=south] at (8.2, -.7) {90.41};
    \node[minimum size=6mm, anchor=south] at (8.2, -.11) {non-rigid};

  \foreach \y [count=\n] in {
      {87.99,85.11,85.91,85.58},
      {69.15,68.69,69.08,69.41},
      {74.10,73.55,74.24,74.63},
      {83.25,81.89,82.50,82.28},
    } {
      \foreach \x [count=\m] in \y {
        \node[fill=purple!\x!yellow, minimum size=6mm,text width=9mm, text=white, align=center] at (2.05*\m,-1.05*\n) {\x};
      }
    }

  \foreach \a [count=\i] in {none,slow,fast,non-rigid} {
    \node[minimum size=6mm, anchor=east] at (1,-1.05*\i) {\a};
  }

  \node[minimum size=6mm] at (4.5, -5.35) {$N_{\emph ft}=10$};
  
\end{tikzpicture}
}
\resizebox{.32\linewidth}{!}{
\begin{tikzpicture}[scale=0.6]

    \node[minimum size=6mm] at (4.5, 1.1) {\textbf{Pretrain}};
    \node[minimum size=6mm, rotate=90, anchor=south] at (-.5, -2.2) {\textbf{Finetune}};

    \node[minimum size=6mm, anchor=south] at (-.1, -.7) {Acc.};
    \node[minimum size=6mm, anchor=south] at (-.1, -.17) {Drift};

    \node[minimum size=6mm, anchor=south] at (2.05, -.6) {\shortstack{none\\94.44}};
    \node[minimum size=6mm, anchor=south] at (4.1, -.6) {\shortstack{slow\\82.34}};
    \node[minimum size=6mm, anchor=south] at (6.15, -.6) {\shortstack{fast\\79.94}};
    \node[minimum size=6mm, anchor=south] at (8.2, -.7) {90.41};
    \node[minimum size=6mm, anchor=south] at (8.2, -.11) {non-rigid};

  \foreach \y [count=\n] in {
      {90.98,88.91,89.45,89.40},
      {72.49,72.62,72.85,73.04},
      {78.02,77.06,77.81,78.18},
      {86.78,85.56,86.44,86.28},
    } {
      \foreach \x [count=\m] in \y {
        \node[fill=purple!\x!yellow, minimum size=6mm,text width=9mm, text=white, align=center] at (2.05*\m,-1.05*\n) {\x};
      }
    }

  \foreach \a [count=\i] in {none,slow,fast,non-rigid} {
    \node[minimum size=6mm, anchor=east] at (1,-1.05*\i) {\a};
  }

  \node[minimum size=6mm] at (4.5, -5.35) {$N_{\emph ft}=15$};
  
\end{tikzpicture}
}
\caption{Drifting type.}
\label{fig:drifting}
\end{subfigure}
\caption{Validations of recordings with different (a) probes, (b) noise levels, and (c) drifting types.}
\label{fig:various_recording}
\blank
\end{figure}

To test the generality of the pre-trained model in different recordings, we perform the validations on the recordings with different characteristics, involving various probes (NP-1.0, NP-2.0, NP-Ultra), noise levels (10uV, 20uV, 30uV) and drifting types (none, slow, fast, non-rigid), as shown in Fig.~\ref{fig:various_recording}. Generally, the differences among different pre-trained models on each fine-tuning recording reduce with the enlarging of $N_{\emph{ft}}$, because the models learn more effectively when more samples are involved during fine-tuning.

For the probes with different geometries, there are two factors affecting the performance of fine-tuning. First, the recording from a more compact probe contains fewer neurons with a fixed neuron density. It is difficult for the model that is pre-trained with the recording containing few neurons to learn general features for various neurons, thus undermining its performance during fine-tuning. Also, because it is easier to distinguish different spikes for the recording with fewer neurons, the model training or fine-tuning for these recordings generally achieves a higher accuracy. Second, the recordings from the same probe have more similarities than those from different probes. Therefore, it is easier to transfer the knowledge when adapting to the recording from the same probe in pre-training, and this consistency can provide slightly higher performance($\sim1$\%) under certain circumstances. However, when more samples are involved during fine-tuning, the pre-trained models can be generally applied to the recordings from different probes without compromising visible accuracy degradation.

As for the various noise levels, the accuracy decreases with the deterioration of the noise involved in the recording. Similarly to cross-probe testing, fine-tuning the model acquired on the recording with the same noise level can result in slight improvements, but still there is no apparent difference even when a few samples are utilized during fine-tuning.

The testing on different drifting types is conducted in four representative patterns, viz., none (no drifting), slow, fast and non-rigid. For the slow and fast drifting, all neurons are moving coherently. All drifting is applied in the z-dimension, i.e., up-and-down. In the slow drifting, the velocity is set to 10 um/s with a drifting range of 30 um, while the fast drifting is configured with a jump with a maximum distance of 15 um every 20 seconds. As for the non-rigid drifting, the neurons are drifting differently, with a velocity and range of 80 um/s and 10 um, respectively. The fine-tuning for the recording without drifting achieves the highest accuracy, followed by non-rigid, fast, and slow. It indicates that the model is more sensitive to the drifting range instead of the velocity, which may be because the neuron positions are learned in the NN and long-distance drifting harasses the utilization of this feature. However, similar to the aforementioned comparisons across different probes and noise levels, the models pre-trained from the recordings with different drifting types have negligible impact on the fine-tuning performance.

In summary, the pre-trained model could be generalized to various probes, noise levels, and drifting types. However, a neuron-rich recording for pre-training is more favourable when a few spikes are provided for fine-tuning.

\subsection{Comparisons with State-of-the-art Spike Sorters}

\begin{table*}[t]
\centering
\renewcommand{\arraystretch}{1.05}
\setlength\tabcolsep{3pt}
\resizebox{\textwidth}{!}{%
\begin{threeparttable}
\caption{Comparisons with The State-of-Art Spike Sorters}
\label{table:others}
\begin{tabular}{llccccccccccccc}
\hline \hline
\multicolumn{2}{l}{\multirow{2}{*}{\textbf{Recording}}}                                           & \multicolumn{2}{c}{Kilosort4~\cite{ks4}$^1$} &  & \multicolumn{2}{c}{Mountainsort5~\cite{mountainsort}$^1$} &  & \multicolumn{2}{c}{HerdingSpikes2~\cite{hs2}$^1$} &  & \multicolumn{4}{c}{\textbf{FaFeSort (this work)}}                  \\ \cline{3-4} \cline{6-7} \cline{9-10} \cline{12-15} 
\multicolumn{2}{l}{}                                                                              & Acc (\%)                 & Time (s)          &  & Acc (\%)                    & Time (s)                    &  & Acc (\%)                & Time (s)                &  & Acc@5$^2$ (\%) & Acc@10$^2$ (\%) & Acc@15$^2$ (\%) & Time (s)      \\ \hline
\multicolumn{2}{l}{\textbf{Base}}                                                                 & \textbf{91.58}           & 346.69            &  & 56.76                       & 1607.58                     &  & 62.54                   & 47.21                   &  & 78.73          & 87.99           & 90.98           & \textbf{1.32} \\ \hline
\multirow{2}{*}{\textbf{Probe}}                                                & NP-2.0           & 89.72                    & 273.43            &  & 50.68                       & 966.36                      &  & 61.73                   & 45.66                   &  & 77.03          & 87.59           & \textbf{90.92}  & \textbf{1.32} \\
                                                                               & NP-Ultra         & 82.03                    & 52.97             &  & 29.12                       & 582.14                      &  & 72.15                   & 40.30                   &  & 95.27          & \textbf{96.73}  & 96.38           & \textbf{1.32} \\ \hline
\multirow{2}{*}{\textbf{Noise}}                                                & 20uV & \textbf{84.25}           & 316.28            &  & 32.36                       & 727.29                      &  & 30.98                   & 40.15                   &  & 66.49          & 78.80           & 82.70           & \textbf{1.32} \\
                                                                               & 30uV & 70.67                    & 268.58            &  & 18.44                       & 436.89                      &  & 17.50                   & 40.73                   &  & 51.57          & 66.56           & \textbf{71.47}  & \textbf{1.32} \\ \hline
\multirow{3}{*}{\textbf{\begin{tabular}[c]{@{}l@{}}Drift\\ Type\end{tabular}}} & slow             & \textbf{85.45}           & 354.25            &  & 39.65                       & 1362.61                     &  & 39.44                   & 45.41                   &  & 59.24          & 69.41           & 73.04           & \textbf{1.32} \\
                                                                               & fast             & \textbf{81.97}           & 366.78            &  & 31.43                       & 1446.23                     &  & 33.08                   & 48.51                   &  & 63.44          & 74.63           & 78.18           & \textbf{1.32} \\
                                                                               & non-rigid        & \textbf{89.68}           & 352.93            &  & 48.51                       & 1354.15                     &  & 54.45                   & 51.66                   &  & 71.49          & 83.25           & 86.78           & \textbf{1.32} \\ \hline \hline
\end{tabular}%
\begin{tablenotes}
    \item[1] The evaluations on these spike sorters are performed with the Spikeinterface package~\cite{spikeinterface}.
    \item[2] We use Acc@$N_{\emph{ft}}$ to denote the accuracy achieved with $N_{\emph{ft}}$ spikes per neuron for finetuning.
\end{tablenotes}
\end{threeparttable}
}
\blank
\end{table*}

We compare our design with the state-of-the-art spike sorters, Kilosort4~\cite{ks4}, Mountainsort5~\cite{mountainsort}, and HerdingSpikes2~\cite{hs2}, in terms of accuracy and elapsed time, as shown in TABLE~\ref{table:others}. For all algorithms, we use the last 50 seconds in the recording for examination.

Compared with all existing spike sorters, our design is significantly faster. Mountainsort5 and HerdingSpikes2 are not accelerated by GPUs. Although they utilized multi-core CPUs for parallel processing, the processing of long-duration recordings was still time-consuming. Kilosort4 was implemented using PyTorch, which is compatible with GPU platforms. However, this algorithm iteratively performed detection and clustering to increase the accuracy of the recording, leading to intensive computations and also significant execution time. Also, the elapsed time of Kilosort4 is dependent on the characteristics of the recording. A more non-ideal recording, e.g., a higher noise level or more complex drifting pattern, would either require more iterations to improve accuracy or lead to accuracy degradation. On the other hand, our design is built with an end-to-end NN and a post-processing scheme that is fully vectorizable and compatible with modern deep learning frameworks like PyTorch. Therefore, our design achieves significant acceleration in execution time compared to all existing spike sorters regardless of the characteristics of the recording.

As for the accuracy of the spike sorters, Kilosort4 achieved the best accuracy among existing spike sorters as it was optimized for the Neuropixels recordings. The accuracy of our design is dependent on the number of spikes utilized for fine-tuning. With $N_{\emph{ft}}=15$, our design achieves comparable or even better performance compared to Kilosort4. Though our method requires manual annotations of the recording, the required number of annotated spikes is relatively small, as we discuss in Section~\ref{sec:experiments}.~2.

\section{Conclusion}

In this letter, we propose E-Sort to perform high-channel-count spike sorting. With transfer learning and parallelizable post-processing, the required number of annotated spikes and processing time are significantly decreased. We examined our design using datasets synthesized with Neuropixels electrode array geometries. Results show that the proposed pre-training-and-fine-tuning framework performs better than training the models from scratch achieving up to 25.68\% better accuracy. Notably, it requires only 44\% number of annotated spikes to achieve an accuracy that is comparable with Kilosort4. Also, the pre-trained model can be applied to recordings from different probes with different noise levels and drifting patterns. Our design only consumes 1.32 seconds for sorting a 50-second recording, which is significantly faster than other state-of-the-art spike sorters.

\ack{This work was supported in part by the Royal Society under grant IEC/NSFC/223067. This work was supported by the Edinburgh International Data Facility (EIDF) and the Data-Driven Innovation Programme at the University of Edinburgh.}

\section*{References}

\bibliographystyle{unsrt}
\bibliography{esort}

\end{document}